\documentclass[conference]{IEEEtran}
\IEEEoverridecommandlockouts

\ifCLASSINFOpdf
   \usepackage[pdftex]{graphicx}
\else
\fi
\usepackage[tight,footnotesize]{subfigure}
\usepackage{url}
\usepackage{hyperref}

\newcommand{\eg}{\emph{e.g., }}
\newcommand{\ie}{\emph{i.e., }}
\newcommand{\etc}{\emph{etc... }}

\newcommand{\cut}[1]{}

\begin{document}
%
\title{Closing the loop of SIEM analysis to\\ Secure Critical Infrastructures 
\thanks{This work has been partially supported by the TENACE PRIN Project (n. 20103P34XC) funded by the Italian Ministry of Education, University and Research.}
\author{\IEEEauthorblockN{Alessia Garofalo\IEEEauthorrefmark{1}, Cesario Di Sarno\IEEEauthorrefmark{1}, Ilaria Matteucci\IEEEauthorrefmark{2},
Marco Vallini\IEEEauthorrefmark{3}, Valerio Formicola\IEEEauthorrefmark{1}}
\IEEEauthorblockA{\IEEEauthorrefmark{1}University of Naples "Parthenope", Department of Engineering, Naples, Italy \\
email: \{alessia.garofalo, cesario.disarno, valerio.formicola\}@uniparthenope.it}
\IEEEauthorblockA{\IEEEauthorrefmark{2}IIT-CNR, Pisa, Italy, email: ilaria.matteucci@iit.cnr.it}
\IEEEauthorblockA{\IEEEauthorrefmark{3}Politecnico di Torino, Dip. di Automatica ed Informatica, Torino, Italy, email: marco.vallini@polito.it}}
}


%


\newtheorem{example}{Example}[section]

\maketitle

\begin{abstract}
Critical Infrastructure Protection is one of the main challenges of last years. Security Information and Event Management (SIEM) systems are widely used for coping with this challenge. However, they currently present several limitations that have to be overcome. In this paper we propose an enhanced SIEM system in which we have introduced novel components to i) enable multiple layer data analysis; ii) resolve conflicts among security policies, and discover unauthorized data paths in such a way to be able to reconfigure network devices. Furthermore, the system is enriched by a Resilient Event Storage that ensures integrity and unforgeability of events stored. 
\end{abstract}


%
\IEEEpeerreviewmaketitle
\thispagestyle{plain}
\pagestyle{plain}

\section{Introduction}\label{sec:intro}
Department of Homeland Security (DHS) defines a Critical Infrastructure in the following way: {\em "Critical infrastructure are the assets, systems, and networks, whether physical or virtual, so vital to the United States that their incapacitation or destruction would have a debilitating effect on security, national economic security, national public health or safety, or any combination thereof"} \cite {ciDefinition}.
Therefore, the protection of these infrastructures must be carefully considered to avoid disasters. For this purpose, DHS has identified sixteen different Critical Infrastructures that need to be monitored and protected \cite {ciReport}.
The study performed by "Industrial Control Systems Cyber Emergency Response Team" (ICS-CERT) \cite{icscert1} highlighted that energy sectors - dams included - are the most attractive targets for cyber-attacks. Recently, the U.S. intelligence agency \cite{damdatabase} traced back a cyber intrusion performed by China government into a database containing sensitive information of the USA government. Specifically, the database stored vulnerabilities of major dams in the United States that can be exploited to perform a future cyber attack against the US electrical power grid. 

Nowadays, as described in McAfee's report \cite{mcafee}, Security Information and Event Management (SIEM) systems are widely used to perform real-time monitoring and control of a Critical Infrastructure. 
SIEM \cite{general_siem} solutions are a combination of the formerly heterogeneous product categories of Security Information Management (SIM) and Security Event Management (SEM).
In particular, SEM systems are focused on the aggregation of data into a manageable amount of information with the help of which security incidents can be dealt with immediately, while SIM primarily focuses on the analysis of historical data in order to improve the long term effectiveness and efficiency of information security infrastructures \cite{william-SIEM}. 
SIEM technology aggregates event data produced by security devices, network infrastructures, systems and applications. 
The primary data source is log data, but SIEM technology can also process other forms of data. Event data is combined with contextual information about users, assets, threats and vulnerabilities. The data is normalized, so that events and contextual information from heterogeneous sources can be correlated and analyzed for specific purposes, such as network security event monitoring, user activity monitoring and compliance reporting. SIEM technology provides real-time security monitoring, historical analysis and other support for incident investigation and compliance reporting.

A weakness of SIEM systems is that they lack several features for Critical Infrastructure protection. In particular, four limits have been identified:
1) SIEMs processing capabilities include collection, aggregation and cross-correlation of heterogeneous sources, typically characterized by different syntax and semantics. However, they cannot process multiple layer data and take into account business process view, service view and physical domain view at the same time. Also, SIEM collectors cannot process data in proximity of the monitored domain in order to limit the amount of information disclosed out of the collection boundaries, \ie typically the part under legislative control of a company.
2) Gartner report \cite{gartner} highlighted as the SIEM lacks context policies that are needed to identify exceptions. Critical Infrastructure monitoring requires instead that many context policies are defined in order to avoid misbehaviour. Also the need for several policies raises additional issues: what happens if some policies generate a conflict? How is it possible to take a decision when two policies are in conflict?
3) Critical Infrastructure monitoring is performed by deploying communication networks that enable the exchange of information between the monitored facilities and the control systems. In order to deny specific connections from external networks towards the internal ones, security policies pose strong limitations to data flows. For instance, an operation of sensor firmware re-writing can only be done from specific hosts in a permitted LAN, where privileged accounts exist and limited access to domain expert profiles is allowed. SIEM systems lack a methodology to identify and control all possible data paths existing in the monitored infrastructures. 
4) SIEMs generate alarms when attack signatures are detected. Alarms are stored along with related events in storage media, \eg databases. Information contained in the alarms can be used for forensic purposes in order to discover the attacker's identity and get details on the attack execution. Thus, alarms' integrity and unforgeability are two requirements that must be ensured to consider them as valid evidence. Today, only few commercial SIEMs ensure these requirements through modules that sign the alarms with classic cryptography algorithms, such as RSA or DES.
This approach is not resilient to attacks against the module that generates signed records. 

In this paper we proposed an enhanced SIEM system that overcomes the limits described above. In particular, the proposed SIEM is designed by integrating: an advanced security information and event collector enabling multiple layer data analysis, namely the Generic Event Translation (GET) framework; the Decision Support System that allows both to resolve the conflict between security policies when it is raised and to analyze/control IT networks. IT networks monitoring and control allow to discover unauthorized data paths and perform automatic re-configuration of network devices; a Resilient Event Storage system that ensures integrity and unforgeability of alarms even in the case of attacks against its components. Finally we analyzed the Hydroelectric Dam Critical Infrastructure. In particular we propose a misuse case that mimics the attack that can be performed by the China government as described above.

\section{Enhanced SIEM Architecture}\label{sec:architecture}
\begin{figure}[htbp]
\centering
\includegraphics[scale=0.50]{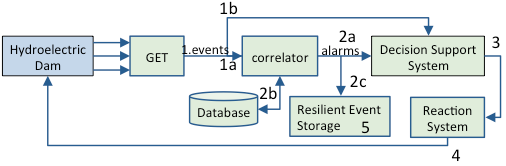}
\caption{Enhanced SIEM Architecture}
\label{encSIEMArc}
\end{figure}

In this section we discuss the conceptual architecture of the enhanced version of the SIEM system we propose. We present it through the block diagram in Fig. \ref{encSIEMArc} where an Hydroelectric Dam is the Critical Infrastructure we want to protect from cyber-attacks. 
These facilities host the information sources, such as, \eg sensors, logical and physical access events, network system, that we monitor through the SIEM. 
Hence, in the following, we firstly describe how an Hydroelectric Dam works and we point out possible security threats. Then we detail the workflow of the proposed architecture.
\subsection{An Hydroelectric Dam}\label{sec:cs} 
\begin{figure}
\centering
\includegraphics[scale=0.15]{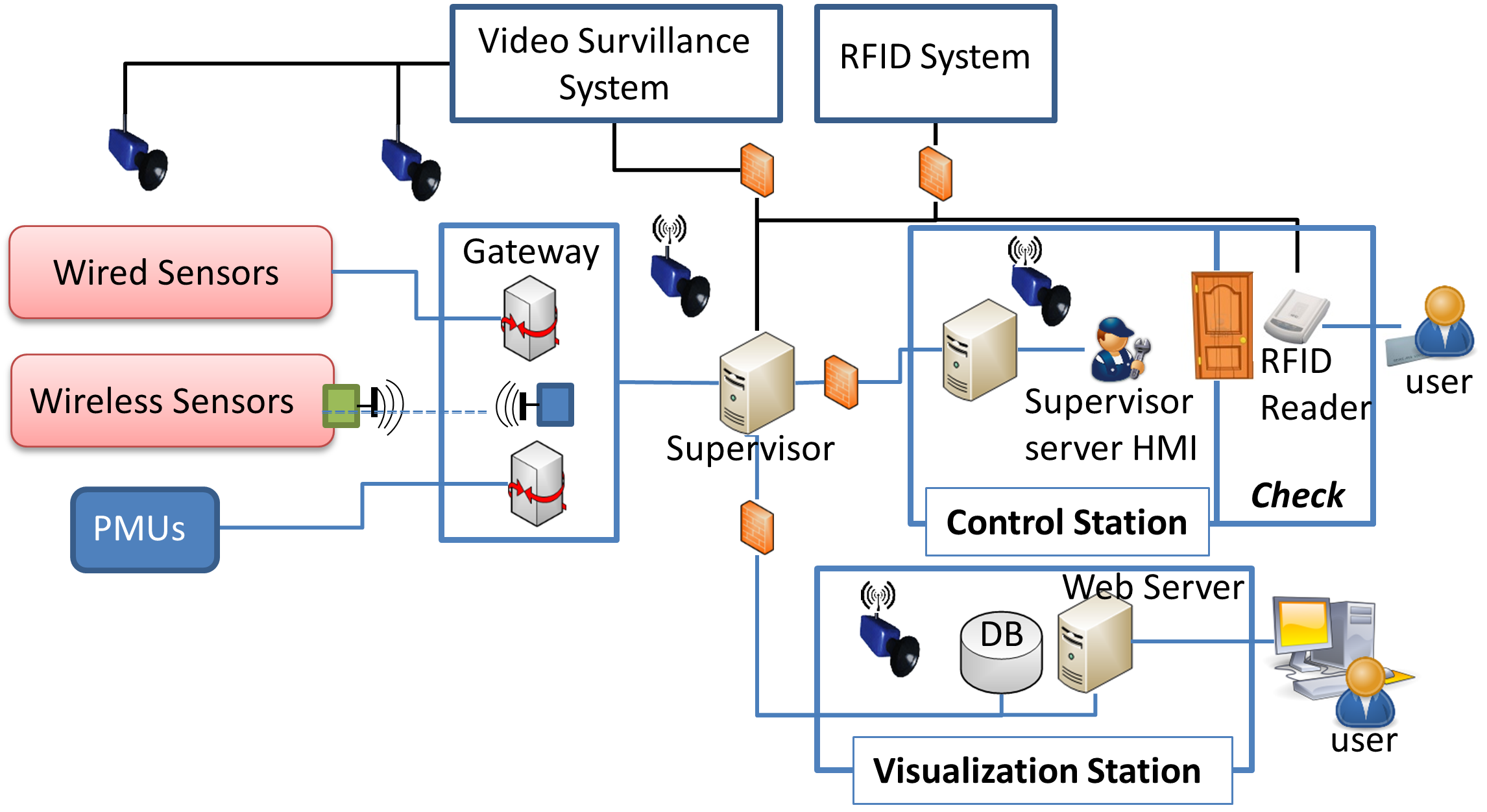}
\caption{IT infrastructure to support dam monitoring and control}
\label{itdam}
\end{figure}

The purpose of a hydroelectric dam is to feed the power grid as any common generator. The simplified process that allows the generation of hydropower is the following:
- the reservoir is fed by a river; - a certain amount of water contained in the reservoir flows through more than one penstocks and move a turbine; - the turbine is linked to an alternator that converts the mechanical energy into electrical energy; - the electrical energy is injected in a transmission line of the power grid. 
The power generated by hydroelectric dam mainly depends on two quantities: the water flow rate $Q$ provided to the turbine through the penstocks and the difference between water level in the reservoir and  water level in the turbine $\Delta h$.
The complete expression used to calculate the power generated by turbine rotation is described from the expression \ref{eq:power}
\begin{equation}
P = \rho*\eta*g*\Delta h*Q
\label{eq:power}
\end{equation}
where: $\rho$ is the water density ($10^3 \frac{Kg}{m^3}$), $\eta$ is the efficiency of the turbine and $g$ is the gravitational constant  ($10 \frac{m}{s^2}$).
If we suppose that $\Delta h$ is a constant in expression \ref{eq:power}, then we obtain that the power is only a function of the water flow rate $Q$. 
With this assumption it is possible to increase the power generated $P$ by increasing $Q$. 
Hence, if the water flow rate increases without control, a state of emergency is generated because this implies an increase of power generated that can lead to a damage of the turbine. Indeed, higher power implies higher number of the turbine rotations. If the number of rotations per minute overcome a fixed threshold, then the turbine is destroyed and/or the electric power generated overcomes the security threshold.
In Fig. \ref{itdam} we show a simplified view of IT systems that are used to monitor and control the dam process. In particular, the 'visualization station' shows data gathered to generate statistics or to monitor a specific process. 
The 'control station' allows to control the process, \eg it is possible to send commands to sensors and actuators. 
Wireless or wired sensors are used to monitor different physical dam parameters in order to assess the safety of the global dam and foresee possible failures or anomalies \cite{Yang:2009:MIS:1726588.1727819}.
Both stations and sensors can be attacked by a malicious user. Indeed, visualization and control stations can be subjected to attacks. For example, due to a misconfiguration the firewall allows an user in the 'visualization station' to re-write the sensor firmware; an unsatisfied employee discovers such wrong configuration in network devices, and so he/she exploits this vulnerability to perform a serious attack to the hydroelectric dam.
Also, wireless sensors can be attacked from insider attackers as well as from outsider ones that try to violate or spoof the communication among sensors in order to obtain some key information useful to violate the dam architecture.
In all these cases, control stations and sensors represent possible points of failure of the architecture and they have to be monitored in order to rapidly identify and solve the occurred security issue.

\subsection{The proposed Architecture}\label{sec:arch}
As depicted in Fig. \ref{encSIEMArc}, the workflow is made of the following steps:\\
\textbf{1.} The events generated by the Hydroelectric Dam are monitored by the GET module that is the security collector software. The purpose of the GET is to generate security events by observing multiple layer data from the sources in the monitored infrastructure. 
The GET translates such events into a common format which is suitable for both the central Correlator engine (1a) and for the Decision Support System (1b). 
\textbf{2.} The Correlator analyzes the GET events to discover known attack signatures, \ie signatures encoded through schematic rules and stored in the rule database of the SIEM (2b). If one of the patterns of security breach is found, then the Correlator generates an alarm and sends it to the Decision Support System (2a) and to the Resilient Event Storage (2c). Generated alarms contain information about the security breach, so they are useful for forensic purposes. 
\textbf{3.} The Decision Support System (based on XACML engine) ensures that the security policies established are not violated and it implements a resolution strategy when two policies are in conflict. Also, the Decision Support System uses a novel modelling approach based on the matricial representation of network device configurations (\ie policies) which allows the computation of the reachability analysis. 
The reachability analysis is an approach useful to discover unauthorized data paths between hosts due to a misconfiguration of network infrastructure. 
\textbf{4.} If the Decision Support System discovers a misconfiguration, it activates the Reaction System that performs a control action on the scenario, \eg a data path discovered and unauthorized is closed by modifying firewalls rules. 
\textbf{5.} In order to use the alarms generated and stored by the SIEM as evidence of cyber-crime, integrity and unforgeability must be guaranteed. The Resilient Event Storage is an intrusion and fault tolerant facility designed to ensure these two requirements even if some components are compromised by an attack.

%
\section{Description of new SIEM Components}
In this section we provide more details about GET module, the Decision Support Systems functionalities, and the Resilient Event Storage system, respectively.
\subsection{Probes and Correlation Process}
\begin{figure}[htbp]
\centering
\includegraphics[scale=0.20]{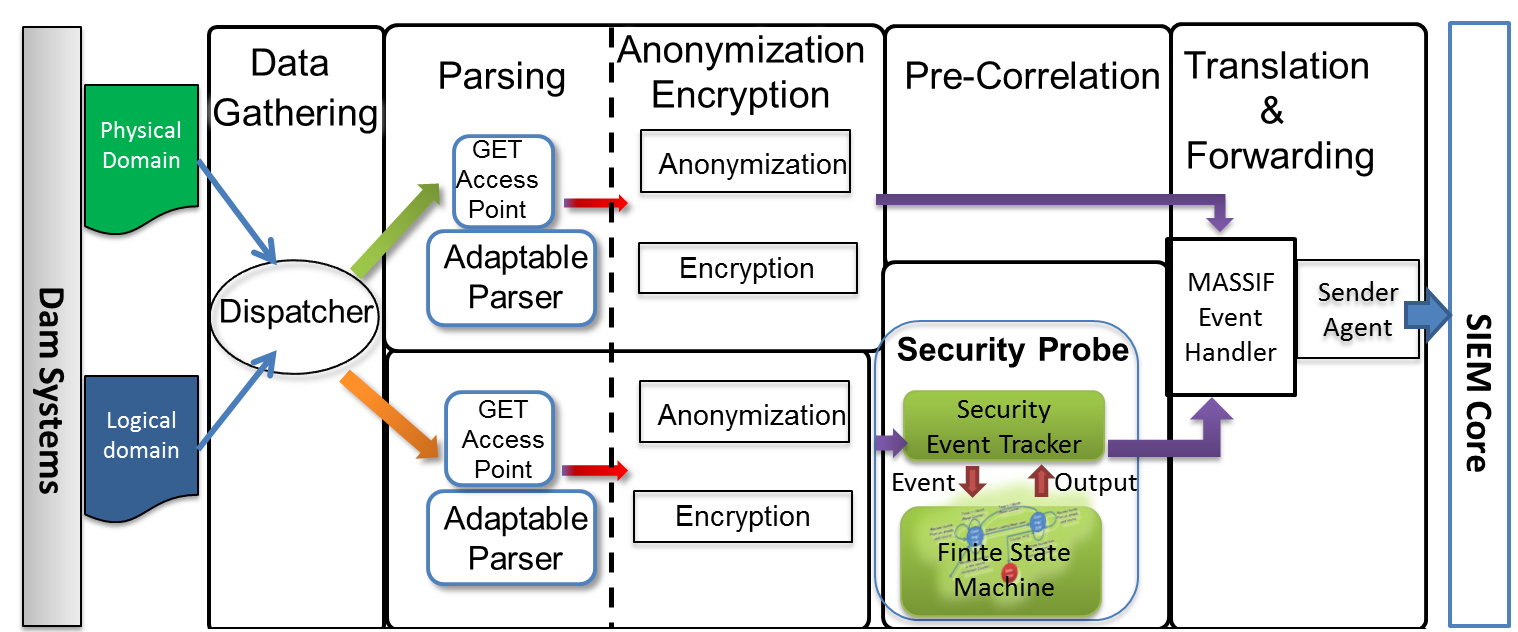}
\caption{Generic Event Translation (GET) framework: architecture}
\label{get}
\end{figure}

\begin{figure}[htbp]
\centering
\includegraphics[scale=0.15]{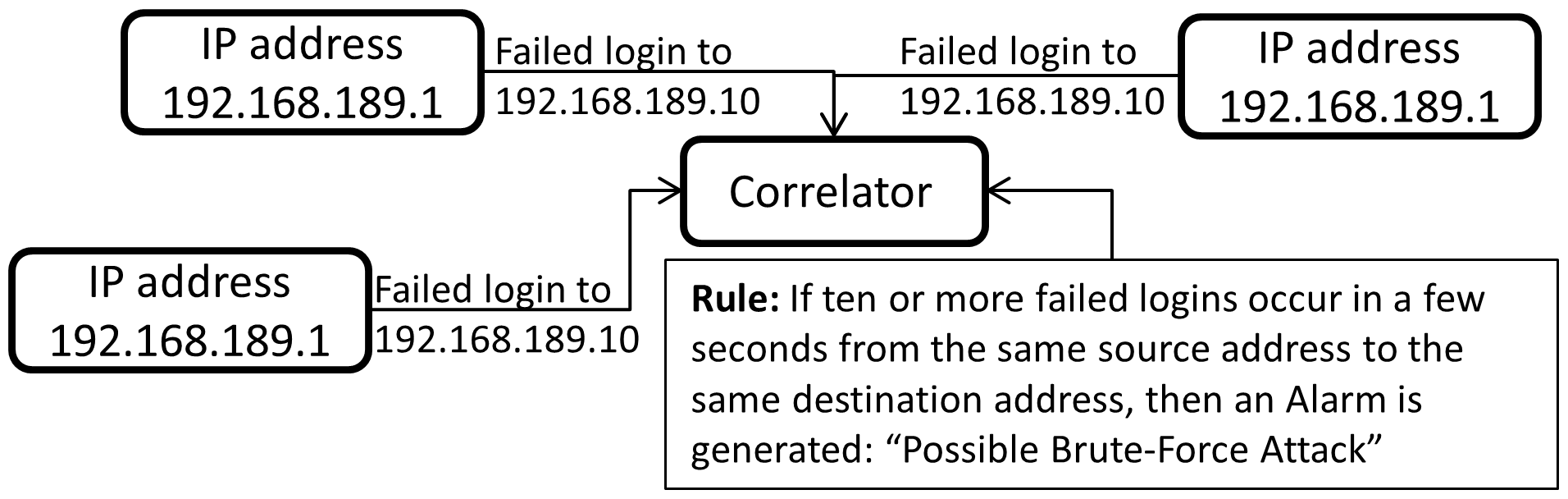}
\caption{Correlation rule example to discover a brute-force attack}
\label{exampleCorrelator}
\end{figure}

The GET framework gathers data from probes deployed within the monitored infrastructure in order to collect security information on specific processes, and generates events when suspicious activities are detected. 
The event format generated by the probes as output of the GET is the same in order to enable processing of the \emph{Correlator} engine. 
In the GET framework the process of excerpting semantically richer information from the logs is realized in a number of sub-tasks including: gathering of raw and heterogeneous data, correlation and abstraction of security relevant facts based on complex analysis~\cite{Formic} \cite{5161659}.
All processes are performed at the edge of the SIEM - \ie in the network of field systems - and are assigned to different modules organized as in Fig.~\ref{get}.
Key components of the GET framework are the \emph{Adaptable Parsers} and the \emph{Security Probes}.
Adaptable Parsers (APs) are high performance data parsers relying on grammar-based compilers. APs are hardcoded parsers obtained by a formal description (grammar) of the data structures and they extract tokens from the data streams given as input to the GET framework. 
Security Probes (SPs) are event pattern detectors based on complex State Machine models. SPs exploit flexibility of statecharts and minimize the effort of implementing all transitions to define security event patterns.
%
In Fig. \ref{encSIEMArc}, \emph{Correlator} engine is a software component that allows to detect specific attacks signatures within events flow received by GET. When an attack signature is matched, the Correlator generates an alarm. The alarm generated contains also information about the events that generated it. Alarm generation through Correlator is performed in order to improve the accuracy of incident diagnosis and allow better response procedures. The Correlator shows few, semantically richer alarms in the face of the huge number of events coming from single sensors.
The well-known attacks signatures are defined through the {\em correlation rules}. In particular, a correlation rule describes a relation between some information contained in the fields of events gathered in order to identify an attack. An example of correlation rule that can be used, for example, to discover a brute-force attack is shown in Fig. \ref{exampleCorrelator}.
In particular the rule establishes that a brute-force attack occurs when many failed logins are performed by same source IP address to the same destination IP address. Thus, if this signature is found within events flow analyzed an alarm is generated and then a reaction can be activated. 

\subsection{Decision Support System}
The Decision Support System exposes two main functionalities: i) a policies conflict resolution strategy that allows to solve conflicts occurring among different XACML-based policies that can be applied at the same time but that allow conflicting actions; ii) a reachability analysis that is able to discover unauthorized network access and allows to re-define network configurations. 
\subsubsection{Policies Conflict Resolution}
\begin{figure}[t]
\centering
\includegraphics[scale=0.4,angle=-90]{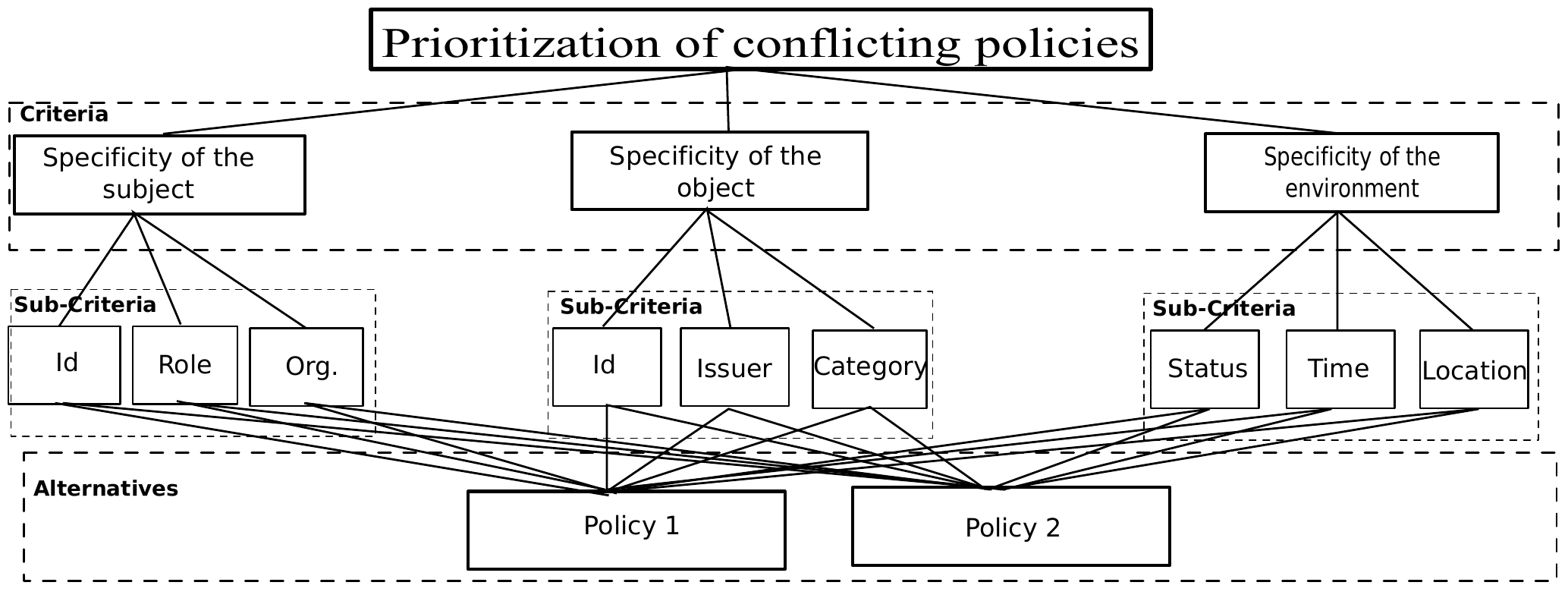}
\caption{AHP Hierarchy for Policy Conflict Resolution~\cite{dpm12,CBMS2013}.}\label{ExampleHierarchy}
\end{figure}

It is based on the Analytical Hierarchy Process (AHP)~\cite{Saaty03,ahp08}, a well-know multi-criteria decision system. 
%
%
The AHP approach requires to
subdivide a complex problem (\ie ranking conflicting policies) into a set of sub-problems, equal in number to the
chosen \emph{criteria}, and then computes the solution (\ie choose the applicable policy) by properly merging all the local solutions for each sub-problem.
In Fig.\ref{ExampleHierarchy} we show a possible instantiation of the AHP hierarchy for policy conflict resolution showed in~\cite{dpm12,CBMS2013}. The goal (the box on top of the hierarchy in Fig.\ref{ExampleHierarchy}) is "select the policy" to  among conflicting ones, \eg Policy 1 and Policy 2 in the boxes at the bottom of the hierarchy.  To solve conflicts we consider as {\it criteria} (second group of boxes starting from the top of the hierarchy) the \emph{specificity} of the elements that constitute a policy: i) \textit{Specificity of the subject}, in which we evaluate the attributes exploited in the two policies to identify the subject, to determine which of the policies define a more specific set of subjects. ii)
\textit{Specificity of the object} in which we evaluate the
             attributes exploited in the two policies to identify the object. 
iii) \textit{Specificity of the environment} in which we evaluate the
              attributes to identify the environment.\\
Furthermore, AHP features the capability to further refine each criterion in sub-criteria, by considering  the attributes that identify each element, \eg for the subject, the Identification Number (ID, the subject Role, the organization the subject belongs to). It is worth noticing that the set of considered attributes depends on the chose scenario.  
%
Once the hierarchy is built, the method performs pairwise comparison, from bottom to top, in order to compute  the relevance, hereafter called \emph{local priority}: i) of each alternative with respect to each sub-criteria, ii) of each sub-criterion with respect to the relative criterion, and finally, iii) of each criterion with respect to the goal. Note that, in case of a criterion without sub-criteria, the local priority of each alternative is computed with respect to the criterion.\\
Comparisons are made through a scale of numbers typical to
AHP (see Table~\ref{scale}) that indicates how many times an alternative
is \emph{more relevant} than another.
\begin{table}[h]
\centering
\medskip
\begin{tabular}{|c|c|c|}
\hline \textbf{Intensity}  &  \textbf{Definition} & \textbf{Explanation}\tabularnewline
\hline
\hline 
1 & Equal & Two elements contribute equally to the objective\tabularnewline
\hline 
3 & Moderate & One element is slightly more relevant than another\tabularnewline
\hline 
5 & Strong & One element is strongly more relevant over another\tabularnewline 
\hline 
7 & Very strong & One element is very strongly more relevant over another\tabularnewline
\hline 
9 & Extreme & One element is extremely more relevant over another\tabularnewline
\hline
\end{tabular}
\smallskip
\caption{Fundamental Scale for AHP\label{scale}}
\vspace{-5mm}
\end{table}

We define an ordering among the attributes related to the same policy element. This ordering expresses how including in a policy a condition on a given attribute contributes to make the policy more
specific. Roughly, attribute {\it a1} of element {\it e} is more {\it specific} than attribute $a2$ of the same element if a condition on this attribute is likely to identify a more homogeneous and/or a smaller set of
entities within {\it e}. For example, in Fig.~\ref{ExampleHierarchy} the subject ID is more specific than the Role.\\
%
%
To calculate local priorities, we perform k 2x2 pairwise comparison matrices, where k is the number of subcriteria (in our case, k=9). Matrices are built according to the presence of the attributes in the policies. Given that $a_{ij}$ is the generic element of one of these matrices:
i) Policy1 and Policy2 contain (or do not contain) attribute A:  then $a_{12} = a_{21} =1$. 
ii) If only Policy1 contains A, than $a_{12} = 9$, and $a_{21} = \frac{1}{9}$.
iii)  If only Policy2 contains A, than $a_{12} = \frac{1}{9}$, and $a_{12} = 9$.
%
%
%
%
%
%
%
Once a comparison matrix has been defined, the local priorities can be computed as the normalized eigenvector associated with the largest eigenvalue of such matrix~\cite{saaty77}.\\ 
Then, moving up in the hierarchy, we quantify how sub-criteria are relevant with respect to the
correspondent criterion.   
Hence, we evaluate how the attributes are relevant to identify the subject, the object and the environment: \eg referring to the subject,
ID is more relevant than
  Role and Organization.  Role and Organization have the same relevance.\\
%
Finally, we quantify how the three criteria are relevant for achieving the goal of
solving conflicts. Without loss of generality, we hypothesize that all the
criteria equally contribute to meet the goal. 
The global priority is calculated according to the following general formula:
\begin{equation}
P^{a_{i}}_{g}  = 
      \sum_{j=1}^{n_1}  \sum_{k=1}^{q(w)}  p^{c_{w}}_{g} \cdot p^{sc^w_k}_{c_{w}} \cdot
      p^{a_{i}}_{sc^w_k}+
      \sum_{j=1}^{n2}   p^{c_{j}}_{g} \cdot p^{a_{i}}_{c_{j}} \label{eq:globprio2}
      \end{equation}
where we have in
mind a hierarchy tree where the leftmost $n1$ criteria have a set of sub-criteria each, while the rightmost $n2$ criteria have no sub-criteria
below them,
and $n1 + n2 = n$ is the number of total criteria;
$q(w)$ is the number of sub-criteria for criterion $c_w$, $p^{c_{w}}_{g}$ is the
local priority of criterion $c_w$ with respect to the goal $g$, $p^{sc^w_k}_{c_{w}}$ is the
local priority of sub-criterion $k$ with respect to criterion $c_w$, and
$p^{a_{i}}_{sc^w_k}$ is the local priority of alternative $a_i$
with respect to sub-criterion $k$  of criterion $c_w$.  $p^{sc^w_k}_{c_{w}}$
and $p^{a_{i}}_{sc^w_k}$ are computed by following the same procedure of the
pairwise comparisons matrices illustrated above.
In this straightforward case, the
pairwise comparison matrix is a 4x4 matrix with all the elements equal to 1,
and the local priorities of the criteria with respect to the goal are simply
0.25 each.\\
It is worth noticing that, in our approach, we do not consider as a decisional criterion the specificity of the action.
This is because we evaluate the action only according to its ID, always present
in a policy. 

\subsubsection{Reachability analysis}\label{sec:DSSPolito}

\begin{figure}[htbp]
\centering
\includegraphics[scale=0.40]{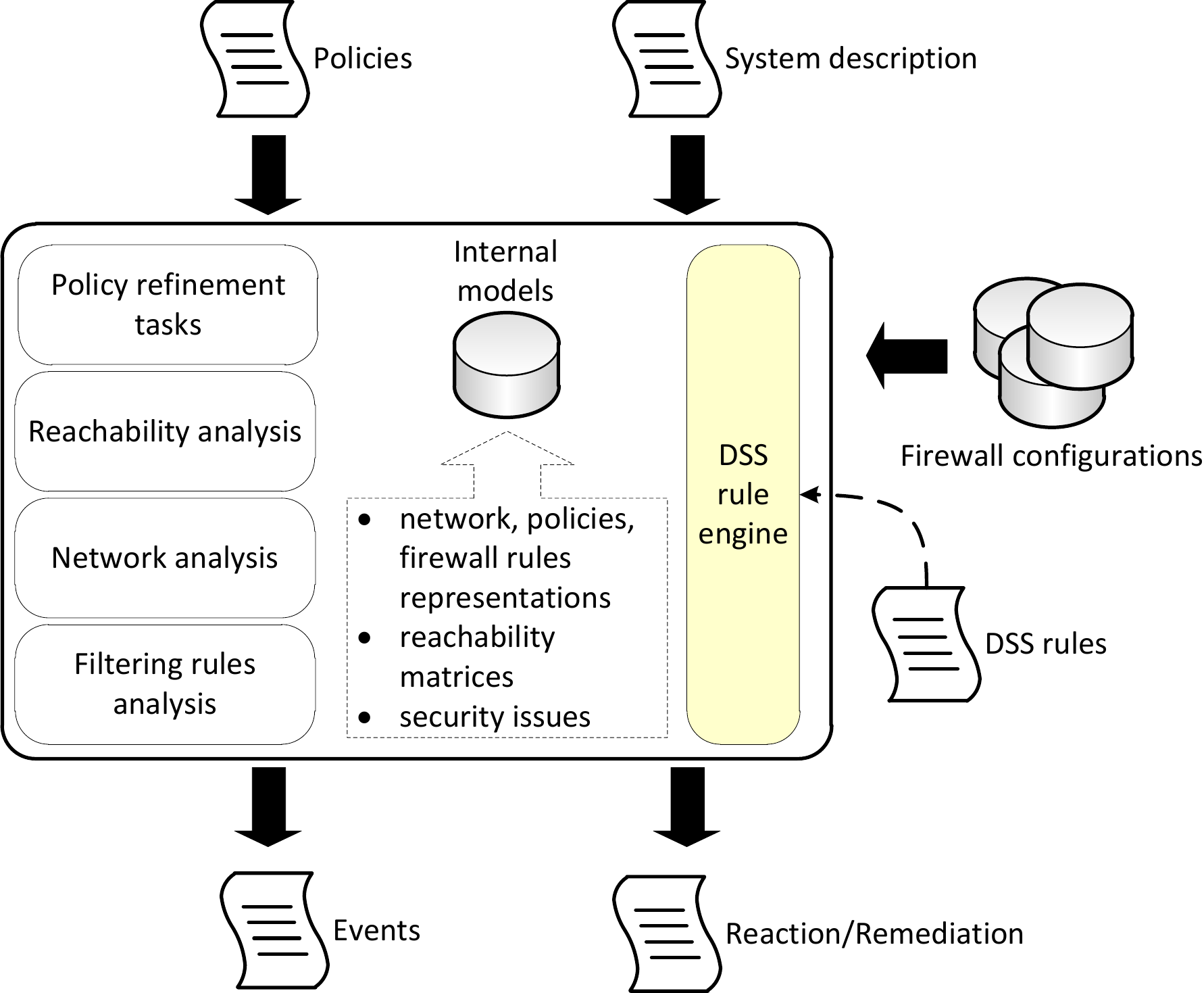}
\caption{Architecture of DSS for reachability analysis}
\label{politoDSS_architecture}
\end{figure}

The objective of DSS for reachability analysis (depicted in Fig. \ref{politoDSS_architecture}) is to discover unauthorized network access. 
An unauthorized network access occurs when firewall rules are modified by non authorized personnel (\eg by an attacker), by authorized personnel (\eg configuration mistakes) or when a policy is not enforceable (\eg no available firewall to enforce the policy).
The main idea to detect these situations is the adoption of network reachability to identify which services are reachable from a set of hosts by traversing devices of a network. 
%
The proposed approach adopts the reachability analysis of the filtering rules comparing the ones derived from policies with the firewall configurations.
However policies are defined by using an abstract language (\eg subject, action, object) and are topology independent (\ie the network topology is not considered during policy authoring). To simplify the policy definition we introduce the action {\em reach} that specifies which network interactions (between subject and object) are authorized. 
This approach simplifies the policy management, \eg the undefined interactions are prohibited as default and policy conflicts are avoided (only some anomalies must be addressed, \eg equivalent rules). 
On the contrary, firewall rules are represented by using a common format (source IP address, source port, destination IP address, destination port, protocol, action) and depend on network topology, \ie where the firewall is placed in the network and which set of hosts protects. 
Therefore policies must be transformed into a concrete format (this operation must be executed for each filtering device of the network) before performing reachability analysis.  
The DSS, by using a set of rules and an inference engine, drives the process to detect unauthorized network access. 
At the beginning, or when policies are modified, DSS starts the refinement process to generate the set of rules for filtering devices. This process is organized by a set of {\em Policy refinement tasks} as depicted in Fig. \ref{politoDSS_architecture}. 
First of all, policies and system description are analysed and a graph-based network topology representation is generated. In particular, during the policy analysis, the anomalies (\eg redundancy) are detected and addressed.  
System description (represented as XML) contains hosts (and related information, \eg IP addresses), capabilities (\eg filtering), services and network topology.
By using {\em Network analysis} the process identifies the set of firewalls to enforce each policy. This task discovers, on the graph-based representation, the network paths (that include at least a firewall) between the subject and the object of a policy. Since the default action of a firewall is to deny all traffic, each firewall contained in a path must be configured to permit the policy traffic. Hence, for each firewall a set of filtering rules is generated to enforce the policies. 
When it does not exist at least a firewall to implement the policy, it is not enforceable. This typically occurs when subject and object belong to the same subnet and their traffic does not traverse any firewall. Therefore any type of traffic between subject and object is permitted, potentially creating a security breach. 
This situation is managed by the DSS, that logs the {\em security issue} and saves it into the {\em internal models} repository.
Once the refinement process is completed, the DSS performs the reachability analysis evaluating the filtering rules, the ones generated by the previous process ({\em generated rules}) and the ones deployed into firewalls ({\em deployed rules}).
This activity is organized in the following phases:\\ 
\noindent I. {\bf Expansion}: For each {\em generated} and {\em deployed} rule, the fields that contain ranges (\eg IP addresses, ports, \etc) are expanded by considering network description (\ie the hosts and services defined into the system description) and creating new rules. Let's consider the rule $r1$ as {\em srcIP:192.168.0.1}, {\em srcPort:*}, {\em dstIP:192.168.10.10}, {\em dstPort:80,443}, {\em proto:TCP}, where the destination IP address refers to a host that offers a web service on ports 80 and 443. The expansion phase transforms $r1$ into rules $r1,1$ and $r1,2$: the former to match the port 80 and the latter to match the port 443. The same approach is followed for the IP addresses expressed as subnets. Before applying expansion operation, {\em deployed} rules are analysed to detect and address anomalies;\\
\noindent II. {\bf Composition}: the objective of this phase is to create the reachability matrices. Each firewall $i$ has two rule sets, one for {\em generated rules} ($R_{g,i}$) and another for {\em deployed rules} ($R_{d,i}$). We introduce the equivalent rule set for the firewall $i$ ($R_{e,i}$) that contains both generated and deployed rules, \ie $R_{e,i} = R_{g,i} \cup R_{d,i}$. Considering the $R_{e,i}$ we create two partitions: the former contains source IP address and port fields ($S_{IP,port}$) and the latter the destination IP address, port and protocol ($D_{IP,port,proto}$). A two-dimensional reachability matrix for a firewall $i$ ($M_i$) has $S_{IP,port}$ as row and $D_{IP,port,proto}$ as column. The composition phase, for each firewall $i$:
	1. creates two matrices: $M_{g,i}$ for {\em generated} rules and $M_{d,i}$ for {\em deployed} rules. Each matrix contains the $S_{IP,port}$ individuals as rows and $D_{IP,port,proto}$ as columns; 
	2. computes $M_{g,i}$: for each rule $r$ of $R_{e,i}$, \ie $r \in R_{e,i}$: if $r$ is part of $R_{g,i}$ rules, \ie $r \in R_{g,i}$, sets $1$ for corresponding row and column, otherwise $0$; 
	3. computes $M_{d,i}$: for each rule $r$ of $R_{e,i}$, \ie $r \in R_{e,i}$: if $r$ is part of $R_{d,i}$ rules, \ie $r \in R_{d,i}$, sets $1$ for corresponding row and column, otherwise $0$;\\	   
\noindent III. {\bf Analysis}: this phase compares reachability property of {\em generated} with {\em deployed} rules. For each firewall $i$, we compute $M_{\rho,i} = M_{g,i} - M_{d,i}$. If $M_{\rho,i} = 0$ (when all the elements are equal to $0$) the reachability for {\em deployed} and {\em generated} rules is the same, \ie no security issues are identified. Otherwise ($M_{\rho,i} \neq 0$) at least an element is equal to $1$ or $-1$. In the first case the corresponding rule is not deployed into the firewall. Therefore, the firewall configuration drops a packet that must be permitted by the policy. This situation is reported as an {\em anomaly}. Otherwise (equal to $-1$) the corresponding rule is enforced by the firewall configuration but it is prohibited by the policy. In this situation, the firewall contains a misconfiguration and the DSS logs it as {\em security issue}.   
%
%
Finally, the DSS evaluating reachability analysis reports detected issues (\ie {\em anomalies}, {\em security issues}) and proposes a {\em remediation}, \ie a list of suggestions on how to modify the firewall rules or where to install a filtering device (\eg personal firewall) to enforce the policy.

\subsection{Resilient Event Storage}
\begin{figure}
\centering
\includegraphics[scale=0.15]{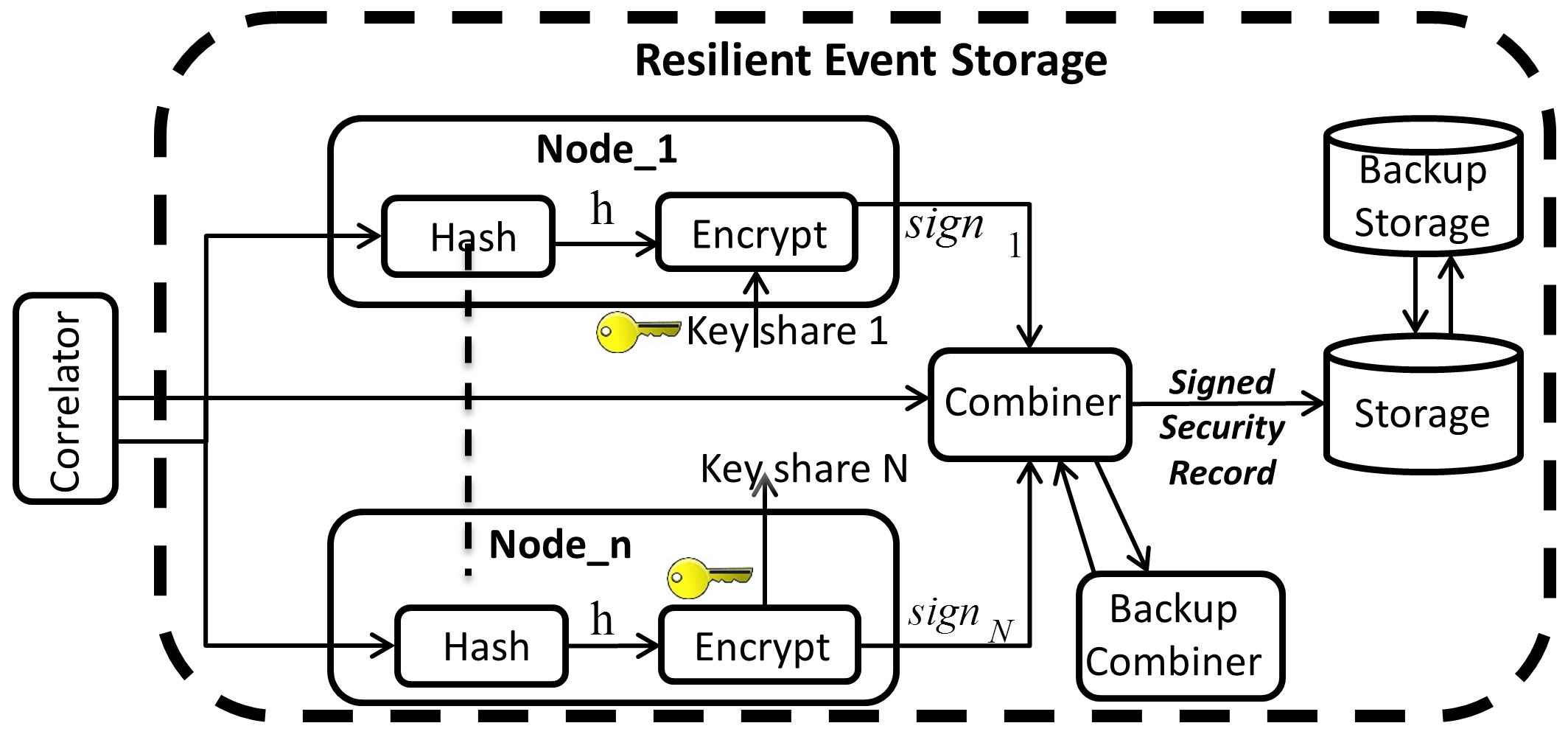}
\caption{Resilient Event Storage Architecture}
\label{res:architecture}
\end{figure}

Resilient Event Storage (RES) system is an infrastructure designed: to be tolerant to faults and intrusions; to generate signed records containing alarms/events related to security breaches; to ensure the integrity and unforgeability of alarms/events stored. In particular, the RES fault and intrusion tolerant capability makes it able to correctly create secure signed records even when some components of the architecture are compromised. The RES conceptual architecture  is shown in Fig. \ref{res:architecture}.
The basic principle is to use more than one secret key. Specifically, the main secret key is one but it is divided in $n$ parts, namely shares, and each share is stored by a different node. This approach can be realized by Shoup threshold cryptography algorithm \cite{shoupAlgorithm}. The most important characteristic of a threshold cryptography algorithm is that the attacker has zero knowledge about the secret key, if less than $(k-1)$ secret key shares are compromised. Threshold cryptography algorithm is characterized by two parameters: $n$ \ie the number of nodes and $k$ \ie the security threshold. The output of a cryptography algorithm and its threshold version are equivalent.
In the RES architecture, a component called Dealer is responsible to generate $n$ secret key shares, $n$ verification key shares and one verification key from a main secret key. This component is not shown in Fig. \ref{res:architecture} because it is used only in the initialization phase. After Dealer generated the keys, it sends each secret key share to each node whereas $n$ verification key shares and the verification key are sent to the Combiner. Each verification key share is used to check the correctness of each signature share generated by each node with its own secret key share. The verification key is used to check the correctness of complete signature generated when the Combiner puts together the signature shares provided by nodes.
Input data to the RES architecture are provided by Correlator (Fig. \ref{encSIEMArc}) because the alarms contain information about a security breach and they need to be stored in secure way. The incoming alarm is sent to all nodes and to the Combiner component. Then, each node computes a hash function of the received alarm. This function returns a digest for this alarm, represented by $h$ in Fig. \ref{res:architecture}. The next step is to encrypt the digest with the secret key share in order to produce a signature share and send it to the Combiner. When the Combiner receives from nodes at least $k$ signature shares for the same alarm it can assemble all partial signatures in order to generate a complete signature. Then the Combiner verifies the complete signature through the verification key. If the verification process fails, the Combiner verifies the correctness of each signature share using the corresponding verification key share. When the node that sent the wrong signature share is discovered it is flagged as corrupted. Next time if new signature shares are available for the same alarm, the Combiner uses the already validated signature shares and the new signature shares to create a new set of $k$ signature shares. Then the Combiner generates a new complete signature and repeats the verification process. If the verification process is successful this time, then the complete signature, the original alert and the identifiers of corrupted nodes are stored in a storage system.
Further details about RES implementation are described in \cite{6375636}.
In order to improve the intrusion and fault tolerance of RES, replication and diversity are employed in the media storage and the Combiner component. 
\section{Conclusion and Future Directions}
In this paper we provided an enhanced SIEM architecture able to cope with cyber security problems that may occur in a Critical Infrastructure. Indeed, starting from a discussion about the limitations of the existing SIEM systems, we proposed a new enhanced SIEM architecture in which we have introduced functionalities for enabling multiple layer data analysis, resolving conflicts among security policies, and discovering unauthorized data paths in such a way to be able to reconfigure network devices.
Also, we provided a sketch of a possible usage of our architecture when a misuse case affects hydroelectric dam. As described in \cite{icscert1}, this Critical Infrastructure is highly exposed to cyber attacks today.
We are currently working to perform an extensive experimental campaign with the purpose of 1) setup the system we proposed in order to integrate all the components and 2) validate the enhanced SIEM proposed in Hydroelectric dam scenario.

\bibliographystyle{IEEEtran}
\bibliography{IEEEfull,biblio}

\end{document}